\documentclass[twocolumn,showpacs,preprintnumbers,amsmath,amssymb,superscriptaddress]{revtex4}

\usepackage{graphicx}% Include figure files
\usepackage{bm}% bold math

\begin{document}
\preprint{arXiv:hep-ph/0609232}
\title{Estimate of {\boldmath ${\cal B}(\bar{B} \to X_s \gamma)$} 
        at {\boldmath ${\cal O}(\alpha_s^2)$}}
\author{M.~Misiak}
\affiliation{Institute of Theoretical Physics, Warsaw University, PL-00-681 Warsaw, Poland}
\affiliation{Theoretical Physics Division, CERN, CH-1211 Geneva 23, Switzerland}
\author{H.~M.~Asatrian}
\affiliation{Yerevan Physics Institute, 375036 Yerevan, Armenia}
\author{K.~Bieri}
\affiliation{Institut f\"ur Theoretische Physik, Universit\"at Bern, CH-3012 Bern, Switzerland}
\author{M.~Czakon}
\affiliation{Institut f\"ur Theoretische Physik und Astrophysik, 
Universit\"at W\"urzburg, D-97074 W\"urzburg, Germany}
\author{A.~Czarnecki}
\affiliation{Department of Physics, University of Alberta, AB~T6G~2J1 Edmonton, Canada}
\author{T.~Ewerth}
\affiliation{Institut f\"ur Theoretische Physik, Universit\"at Bern, CH-3012 Bern, Switzerland}
\author{A.~Ferroglia}
\affiliation{Physikalisches Institut, Albert-Ludwigs-Universtit\"at, D-79104 Freiburg, Germany}
\author{P.~Gambino}
\affiliation{INFN, Torino \& Dipartimento di Fisica Teorica, 
                   Universit\`a di Torino, I-10125 Torino, Italy}
\author{M.~Gorbahn}
\affiliation{Institut f\"ur Theoretische Teilchenphysik, Universit\"at Karlsruhe (TH), D-76128 Karlsruhe, Germany}
\author{C.~Greub}
\affiliation{Institut f\"ur Theoretische Physik, Universit\"at Bern, CH-3012 Bern, Switzerland}
\author{U.~Haisch}
\affiliation{Institut f\"ur Theoretische Physik, Universit\"at Z\"urich, CH-8057 Z\"urich, Switzerland}
\author{A.~Hovhannisyan}
\affiliation{Yerevan Physics Institute, 375036 Yerevan, Armenia}
\author{T.~Hurth}
\affiliation{Theoretical Physics Division, CERN, CH-1211 Geneva 23, Switzerland}
\affiliation{SLAC, Stanford University, Stanford, CA 94309, USA}
\author{A.~Mitov}
\affiliation{Deutsches Elektronen-Synchrotron DESY, D-15738 Zeuthen, Germany}
\author{V.~Poghosyan}
\affiliation{Yerevan Physics Institute, 375036 Yerevan, Armenia}
\author{M.~\'Slusarczyk}
\affiliation{Department of Physics, University of Alberta, AB~T6G~2J1 Edmonton, Canada}
\author{M.~Steinhauser}
\affiliation{Institut f\"ur Theoretische Teilchenphysik, Universit\"at Karlsruhe (TH), D-76128 Karlsruhe, Germany}
\date{September 18, 2006}
\begin{abstract}
  Combining our results for various ${\cal O}(\alpha_s^2)$ corrections to
  the weak radiative $B$-meson decay, we are able to present the first
  estimate of the branching ratio at the next-to-next-to-leading order in QCD.
  We find ${\cal B}({\bar B}\to X_s\gamma) = (3.15 \pm 0.23) \times 10^{-4}$ 
  for $E_{\gamma} > 1.6\;$GeV in the ${\bar B}$-meson rest frame. The four
  types of uncertainties: nonperturbative (5\%), parametric (3\%),
  higher-order (3\%) and $m_c$-interpolation ambiguity (3\%) have been
  added in quadrature to obtain the total error.
\end{abstract}

\pacs{12.38.Bx, 13.20.He}
                             
\maketitle

The inclusive radiative $B$-meson decay provides important constraints on the
minimal supersymmetric standard model and many other theories of new physics
at the electroweak scale. The power of such constraints depends on the
accuracy of both the experiments and the standard model (SM) calculations. The
latest measurements by Belle and BABAR are reported in
Refs.~\cite{Koppenburg:2004fz,Aubert:2005cu}. The world average performed by
the Heavy Flavor Averaging Group~\cite{unknown:2006bi} for~ $E_{\gamma} >
1.6\;{\rm GeV}$ reads
\begin{equation} \label{hfag}
{\cal B}(\bar{B} \to X_s \gamma)~ =~ 
\left(3.55\pm 0.24{\;}^{+0.09}_{-0.10}\pm0.03\right)\times 10^{-4}.
\end{equation}
The combined error in the above result is of the same size as the expected
${\cal O}(\alpha_s^2)$ next-to-next-to-leading order (NNLO) QCD corrections to
the perturbative decay width $\Gamma(b \to X_s^{\rm parton}\gamma)$, and
larger than the known nonperturbative corrections to the relation
$\Gamma(\bar{B} \to X_s \gamma) ~\simeq~ \Gamma(b \to X_s^{\rm parton}\gamma)$
%
%\cite{nonpert1,Buchalla:1997ky,Lee:2006wn}.
\cite{nonpert1}--\cite{Lee:2006wn}.
Thus, calculating the SM prediction for the $b$-quark decay rate at the NNLO
% in QCD 
is necessary for taking full advantage of the measurements.

Evaluating the ${\cal O}(\alpha_s^2)$ corrections to ${\cal B}(b \to X_s^{\rm
  parton} \gamma)$ is a very involved task because hundreds of three-loop
on-shell and thousands of four-loop tadpole Feynman diagrams need to be
computed. In a series of papers 
%
% \cite{Bieri:2003ue, Misiak:2004ew, Gorbahn:2004my,
%  Melnikov:2005bx, Blokland:2005uk, Asatrian:2006ph,
%  Czakon:2006ss, Misiak:2006ab},
\cite{Bieri:2003ue}--\cite{Misiak:2006ab},
we have presented partial contributions to this enterprise. The purpose of the
present Letter is to combine all the existing results and obtain the first
estimate of the branching ratio at the NNLO. We call it an estimate rather
than a prediction because some of the numerically important contributions have
been found using an interpolation in the charm quark mass, which introduces
uncertainties that are difficult to quantify.
\begin{figure}[t]
\begin{center}
\includegraphics[width=4cm,angle=0]{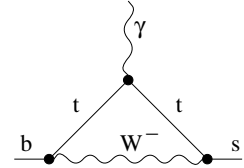}
\caption{\sf Sample LO diagram for the $b\to s\gamma$ transition.\label{fig:LOdiag}} 
\end{center} 
\end{figure}

Let us begin with recalling that the leading-order (LO) contribution to the
considered decay originates from one-loop diagrams in the SM. An example of
such a diagram is shown in Fig.~\ref{fig:LOdiag}. Dressing this diagram with
one or two virtual gluons gives examples of diagrams that one encounters at
the next-to-leading order (NLO) and the NNLO. In addition, one should include
diagrams describing the bremsstrahlung of gluons and light quarks. 

An additional difficulty in the analysis of the considered decay is the
presence of large logarithms $(\alpha_s \ln M_W^2/m_b^2)^n$ that should be
resummed at each order of the perturbation series in $\alpha_s$. To
do so, one employs a low-energy effective theory that arises
after decoupling the top quark and the heavy electroweak bosons. Weak
interaction vertices (operators) in this theory are either of dipole type
($\bar s \sigma^{\mu\nu} b F_{\mu\nu}$,~ $\bar s \sigma^{\mu\nu} T^a b
G^a_{\mu\nu}$) or contain four quarks ($[\bar s \Gamma b][\bar q \Gamma' q]$).
Coupling constants at these vertices (Wilson coefficients) are first evaluated
at the electroweak renormalization scale $\mu_0 \sim m_t, M_W$ by solving the
so-called~ {\em matching}~ conditions.  Next, they are evolved down to the
low-energy scale $\mu_b \sim m_b$ according to the effective theory
renormalization group equations (RGE).  The RGE are governed by the operator~
{\em mixing}~ under renormalization.  Finally, one computes the~ 
{\em matrix elements}~ of the operators, which in our case amounts 
to calculating on-shell diagrams with single insertions of the effective theory 
vertices.

A summary of the $\bar{B} \to X_s \gamma$ calculation status before the beginning of
our project can be found, e.g., in Ref.~\cite{Hurth:2003vb}.  At
the NNLO level, the dipole and the four-quark operators need to be matched up
to three and two loops, respectively. Renormalization constants up to four
loops must be found for $b \to s \gamma$ and $b \to s g$ diagrams with
four-quark operator insertions, while three-loop mixing is sufficient in the
remaining cases. Two-loop matrix elements of the dipole operators and
three-loop matrix elements of the four-quark operators must be evaluated in
the last step.

Three-loop dipole operator matching was found in Ref.~\cite{Misiak:2004ew}.
The necessary three-loop mixing was calculated in Ref.~\cite{Gorbahn:2004my}.
The four-loop mixing was evaluated in Ref.~\cite{Czakon:2006ss}.
Two-loop matrix element of the photonic dipole operator together
with the corresponding bremsstrahlung was found in
Refs.~\cite{Blokland:2005uk,Melnikov:2005bx} and recently confirmed in
Ref.~\cite{Asatrian:2006ph}. Three-loop matrix elements of
the four-quark operators were found in Ref.~\cite{Bieri:2003ue} within the
so-called large-$\beta_0$ approximation. A calculation that goes beyond this
approximation by employing an interpolation in the charm quark mass $m_c$ has
just been completed in Ref.~\cite{Misiak:2006ab}.

\begin{figure}[t]
\begin{center}
\includegraphics[width=82mm,angle=0]{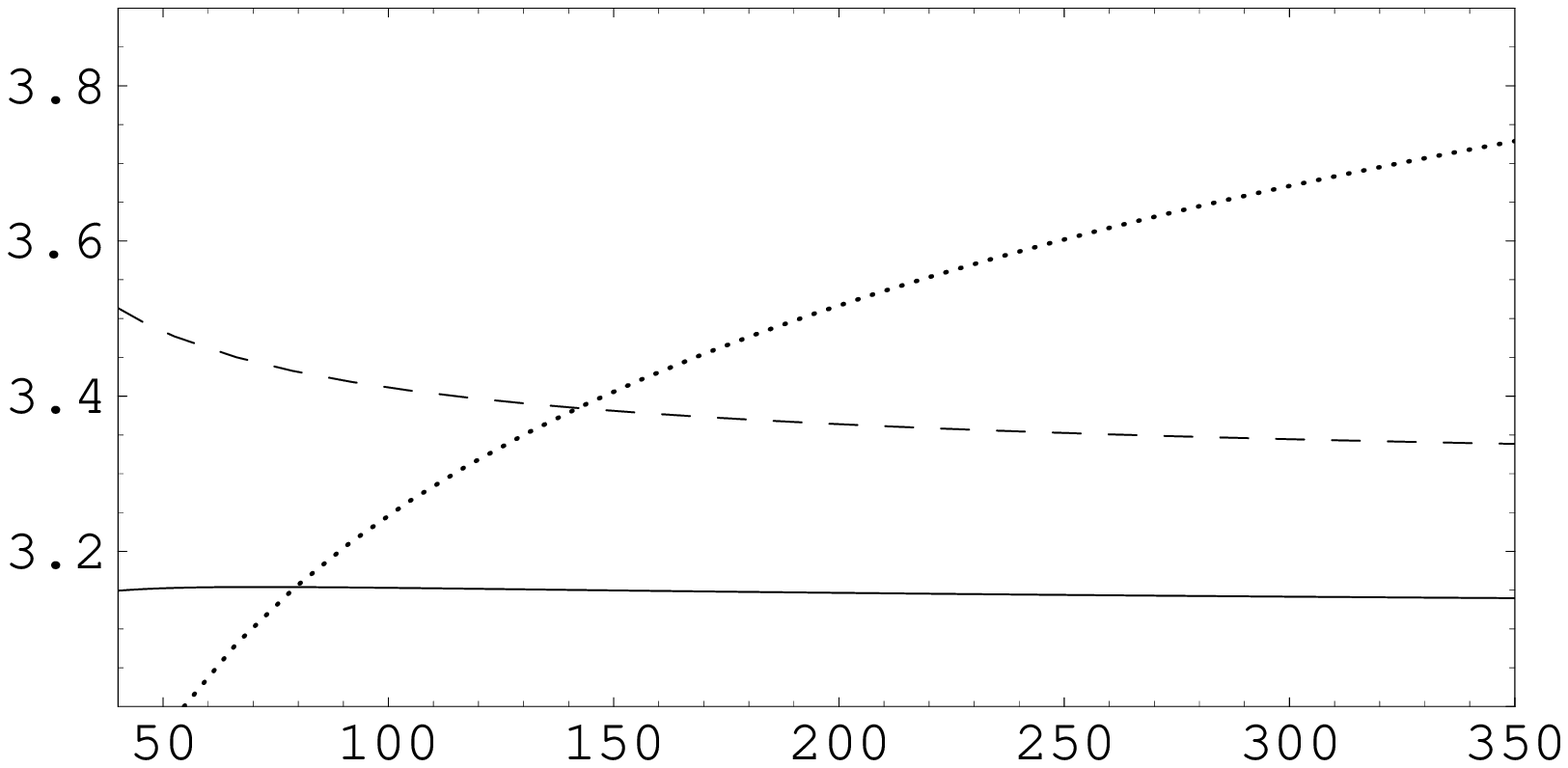}\\[4mm]
\includegraphics[width=82mm,angle=0]{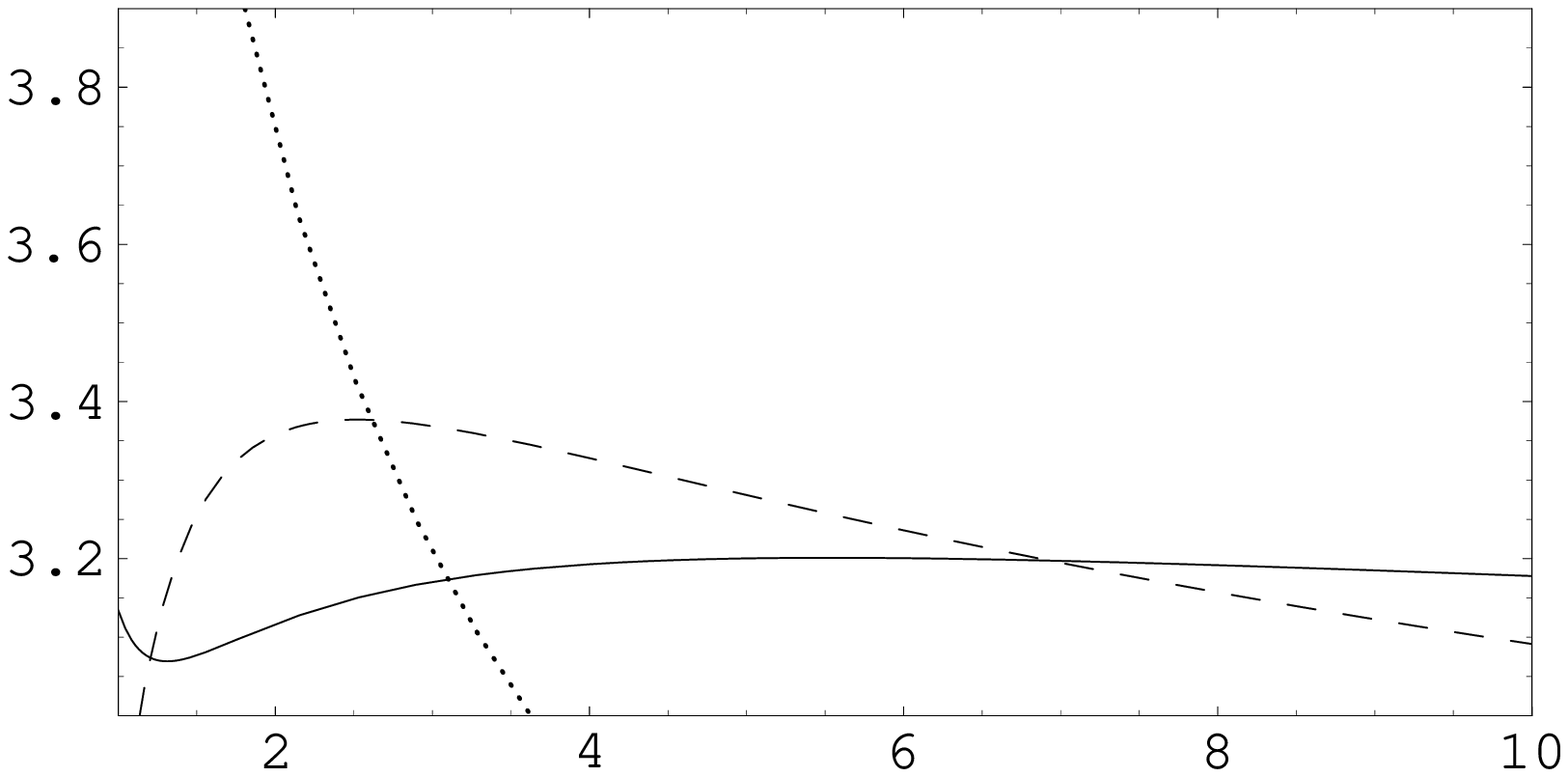}\\[4mm]
\includegraphics[width=82mm,angle=0]{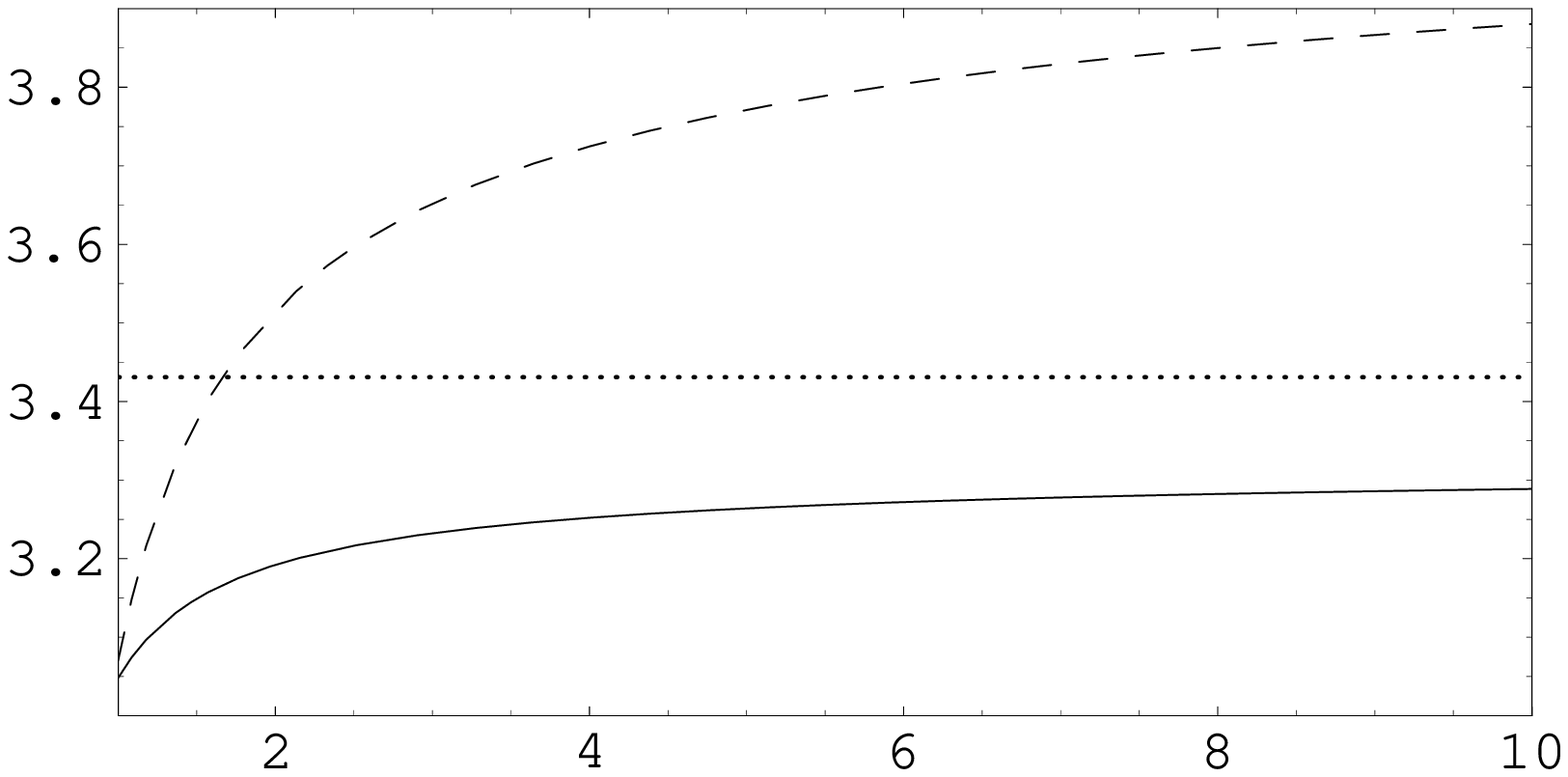}\\[-129mm]
${\cal B}\times 10^4$\hspace{55mm}\mbox{\ } \\[27.5mm]
$\mu_0\;$[GeV]\\[9.5mm]
${\cal B}\times 10^4$\hspace{55mm}\mbox{\ } \\[28mm]
$\mu_b\;$[GeV] \\[9.5mm]
${\cal B}\times 10^4$\hspace{55mm}\mbox{\ } \\[27.5mm]
$\mu_c\;$[GeV]\\[4mm]
\caption{\sf Renormalization scale dependence of ${\cal B}(\bar{B} \to X_s \gamma)$
  in units $10^{-4}$ at the LO (dotted lines), NLO (dashed lines) and NNLO (solid
  lines). The plots describe subsequently the dependence on the matching scale
  $\mu_0$, the low-energy scale $\mu_b$, and the charm mass renormalization
  scale $\mu_c$. \label{fig:mudep}} 
\end{center} 
\end{figure}

With all these results at hand, we are ready to present the first estimate of
the $\bar{B} \to X_s \gamma$ branching ratio at ${\cal O}(\alpha_s^2)$. It
reads~\footnote{
The small ($-0.35\%$) correction from the four-loop $b \to s g$
     mixing diagrams is not included in our numerical results.}
\begin{equation} \label{final}
{\cal B}({\bar B}\to X_s\gamma) = (3.15 \pm 0.23) \times 10^{-4},
\end{equation}
for $E_{\gamma} > 1.6\;$GeV in the ${\bar B}$-meson rest frame. The four types
of uncertainties: nonperturbative (5\%), parametric (3\%), higher-order
(3\%) and $m_c$-interpolation ambiguity (3\%) have been added in quadrature 
in Eq.~(\ref{final}).
% to obtain the %total error. 

The central value in Eq.~(\ref{final}) was obtained for $\mu_0 = 160\;$GeV,
$\mu_b = 2.5\;$GeV and $\mu_c = 1.5\;$GeV. The latter quantity stands for the
charm mass $\overline{\rm MS}$ renormalization scale that is allowed to be
different from $\mu_b$. The branching ratio dependence on each of the three
scales is shown in Fig.~\ref{fig:mudep}. Once one of them is varied, the
remaining two are fixed at the values that have been mentioned above. The
reduction of the renormalization scale dependence at the NNLO is clearly seen.
The most pronounced effect occurs for $\mu_c$ that was the main source of
uncertainty at the NLO. (The LO results are $m_c$- and thus 
$\mu_c$-independent.) The current uncertainty of $\pm 3\%$ due to
higher-order $\left[{\cal O}(\alpha_s^3)\right]$ effects is estimated
from the NNLO curves in Fig.~\ref{fig:mudep}.

The reference value of $\mu_b = 2.5\;$GeV that we have chosen is roughly twice
smaller than in the previous LO and NLO analyses.  Given the stability of the
NNLO result for large values of $\mu_b$, we do not underestimate any
uncertainty from that region. Furthermore, because the center-of-mass energy
$m_B \simeq 5.3\;$GeV gets distributed among various partons, the reference
value of $\mu_b = 2.5\;$GeV seems reasonable.  Lower values of $\mu_b$ have an
advantage of making $\mu_c$-stabilization more efficient because the
NNLO logarithm that compensates $\mu_c$-dependence of the NLO
amplitude comes multiplied by $\alpha_s(\mu_b)$.

The $\pm 3\%$ uncertainty that is assigned to the $m_c$-interpolation
ambiguity has been estimated studying by how much the NNLO branching ratio
depends on various interpolation assumptions. More details on this point and
other elements of the phenomenological analysis (including the 
input parameters) can be found in Ref.~\cite{Misiak:2006ab}.

As far as the parametric uncertainties are concerned, the dominant ones come
from $\alpha_s(M_Z)$ ($\pm 2.0\%$) and the measured semileptonic
branching ratio ${\cal B}(\bar{B} \to X_c e \bar\nu)$ ($\pm 1.6\%$) to
which we normalize.  The third-to-largest uncertainty ($\pm 1.1\%$) is
due to the correlated errors in $m_c(m_c)$ and the semileptonic
phase-space factor 
\begin{equation} \label{phase1}
C = \left| \frac{V_{ub}}{V_{cb}} \right|^2 
\frac{\Gamma[\bar{B} \to X_c e \bar{\nu}]}{\Gamma[\bar{B} \to X_u e \bar{\nu}]}.
\end{equation}
The factor $C$ has been determined in Ref.~\cite{Bauer:2004ve} together with
$m_c(m_c)$ from a global fit to the semileptonic data. If the normalization to
${\cal B}(\bar{B} \to X_c e \bar\nu)$ was not applied in the $\bar{B} \to X_s
\gamma$ calculation, the error due to $m_c(m_c)$ would amount to $\pm 2.8\%$.
At the same time, one would need to take into account uncertainties in $m_b^5$
and the Cabibbo-Kobayashi-Maskawa factor $|V_{ts}^\star V_{tb}|^2$,
each of which exceeds $\pm 3\%$.

The nonperturbative uncertainty in Eq.~(\ref{final}) is due to matrix
elements of the four-quark operators in the presence of one gluon 
that is not soft ($Q^2 \sim m_b^2, m_b \Lambda$, where $\Lambda \sim
  \Lambda_{\scriptscriptstyle\rm QCD}$). Unknown nonperturbative
corrections to them scale like
  $\alpha_s \Lambda/m_b$~ in the limit~ $m_c \ll m_b/2$~
and like
  $\alpha_s \Lambda^2/m_c^2$~ in the limit~ $m_c \gg m_b/2$.
Because $m_c < m_b/2$ in reality, $\alpha_s \Lambda/m_b$~ should be considered as
  the quantity that sets the size of such effects.  Consequently, a $\pm 5\%$
  nonperturbative uncertainty has been assigned to the result in
  Eq.~(\ref{final}).  This is the dominant uncertainty at present. Thus, a
  detailed analysis of such effects would be more than welcome. So far, no published
  results on this issue exist.  Even lacking a trustworthy method for
  calculating such effects, it might be possible to put rough upper bounds on
  them that could supersede the current guess-estimate of~ $\pm 5\%$.
  Nonperturbative corrections to inclusive $\bar B \to X_{d,s}\gamma$ decays that
  scale like $\Lambda/m_b$ may arise when the $b$-quark annihilation vertex
  does not coincide with the hard photon emission vertex; see, e.g.,
  Ref.~\cite{Lee:2006wn} or comments on $\bar B \to X_d \gamma$ 
  in Sec.~2 of Ref.~\cite{Buchalla:1997ky}.
  
  The NNLO central value in Eq.~(\ref{final}) differs from some of the
  previous NLO predictions by between 1 and 2 error bars of the NLO results.
  Because those error bars were obtained by adding various theoretical
  uncertainties in quadrature, such a shift is not improbable, similarly to
  shifts by less than $2\sigma$ in experimental results. The shift from the
  NLO to the NNLO level diminishes with lowering the value of $\mu_c$, which
  has motivated us to use the relatively low $\mu_c = 1.5\;$GeV as a reference
  value here.
  
  The NNLO results turn out to be only marginally dependent on whether one
  follows (or not) the approach of Ref.~\cite{Gambino:2001ew} where the
  top-quark contribution to the decay amplitude was calculated separately and
  rescaled by quark mass ratios to improve convergence of the perturbation
  series. Although the top contribution alone indeed behaves better also at
  the NNLO level when such an approach is used, the charm quark contribution
  (to which no rescaling has been applied in Ref.~\cite{Gambino:2001ew}) does
  not turn out to be particularly stable beyond the NLO. Consequently, in
  the derivation of Eq.~(\ref{final}) and Fig.~\ref{fig:mudep}, we have used
  the simpler method of treating charm and top sectors together.
  
  Our result in Eq.~(\ref{final}) has been obtained under the assumption that
  the photonic dipole operator contribution to the integrated
  $E_{\gamma}$ spectrum below $1.6\;$GeV is well approximated by a fixed-order
  perturbative calculation (see Note added). For lower
  values of the photon energy cut, the following numerical fit can be used:
\begin{equation} \label{fit}
\left( \frac{{\cal B}(E_\gamma > E_0)}{
             {\cal B}(E_\gamma > 1.6\;{\rm GeV})} \right)_{\hspace{-2mm}
\begin{array}{c} \ \\[-2mm] {\scriptstyle\rm fixed}\\[-2mm]
                            {\scriptstyle\rm order} \end{array}} 
\simeq 1 + 0.15 x - 0.14 x^2,
\end{equation}
where $x = 1 - E_0/(1.6\,$GeV). This formula coincides with our NNLO results
up to $\pm 0.1\%$ for $E_0 \in [1.0,\;1.6]\;$GeV. The error is
practically $E_0$-independent in this range.

\begin{figure}[t]
\includegraphics[width=8cm,angle=0]{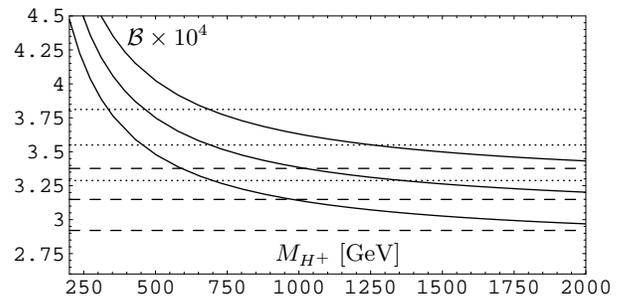}\\[-37mm]
${\cal B}\times 10^4$\hspace{37mm}\mbox{\ }\\[25mm]
\hspace*{6mm} $M_{H^+}\;$[GeV]\\[3mm]
\caption{\sf ${\cal B}(\bar{B} \to X_s \gamma)$ as a function of the charged
  Higgs boson mass in the THDM~II for $\tan\beta=2$ (solid lines). The dashed
  and dotted lines show the SM and experimental results, respectively (see the
  text). \label{fig:MHc}} 
\end{figure}

In the remainder of this Letter, we shall update the $\bar B \to X_s
  \gamma$ constraints on the charged Higgs boson mass in the
  two-Higgs-doublet-model~II (THDM~II) \cite{Abbott:1979dt}. The solid lines
  in Fig.~\ref{fig:MHc} show the dependence of ${\cal B}(\bar{B} \to X_s
  \gamma)$ on this mass when the ratio of the two vacuum expectation values,
  $\tan\beta$, is equal to 2. The dashed and dotted lines show the SM (NNLO)
  and the experimental results, respectively. In each case, the middle line is
  the central value, while the other two lines indicate uncertainties that one
  obtains by adding all the errors in quadrature. 
  
  In our THDM calculation, matching of the Wilson coefficients at the
  electroweak scale is complete up to the NLO~\cite{Ciuchini:1997xe}, but the
  NNLO terms contain only the SM contributions (the THDM ones remain
  unknown). In consequence, the higher-order uncertainty becomes somewhat larger.
  This effect is estimated by varying the matching scale $\mu_0$ from half to
  twice its central value. It does not exceed $\pm 1\%$ for the $M_{H^+}$ range in
  Fig.~\ref{fig:MHc}.

  Even though the experimental result is above the SM one,
  the lower bound on $M_{H^+}$ for a generic value of $\tan\beta$ remains
  stronger than what one can derive from any other currently available
  measurement. If all the uncertainties are treated as Gaussian and combined 
  in quadrature, the 95\%~(99\%)~CL bound amounts to around 295~(230)$\,$GeV. 
  It is found for $\tan\beta \to \infty$ but stays practically 
  constant down to $\tan\beta \simeq 2$. For smaller $\tan\beta$, the branching 
  ratio and the bound on $M_{H^+}$ increase.

\begin{figure}[t]
\includegraphics[width=8cm,angle=0]{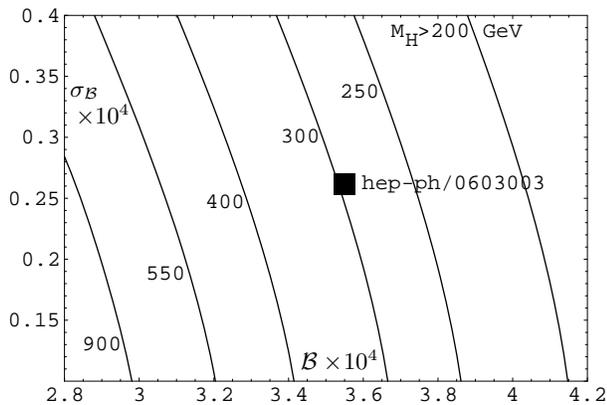}\\[-45mm]
$\sigma_{\cal B}$\hspace{58mm}\mbox{\ }\\
$\times 10^4$\hspace{53mm}\mbox{\ }\\[29mm]
${\cal B}\times\! 10^4$\hspace{-1cm}\mbox{\ }\\[4mm]
\caption{\sf The 95\%~CL lower bound on $M_{H^+}$ as a function of the
  experimental central value (horizontal axis) and error (vertical
  axis). The experimental result from Eq.~(\ref{hfag}) is indicated by the black 
  square. The contour lines represent values that lead to the same bound.
\label{fig:bounds}} 
\end{figure}

The contour plot in Fig.~\ref{fig:bounds} shows the dependence of the
$M_{H^+}$ bound on the experimental central value and error. The current
experimental result (\ref{hfag}) is indicated by the black square.
Consequences of the future upgrades in the measurements will easily be read
out from the plot, so long as no progress on the theoretical side is made. Of
course, the derived bounds should be considered illustrative only because they
depend very much on the theory uncertainties that have no statistical
interpretation. 

To conclude, we have provided the first estimate of ${\cal B}(\bar B \to X_s
\gamma)$ at ${\cal O}(\alpha_s^2)$. The inclusion of the NNLO QCD corrections
leads to a significant suppression of the branching ratio renormalization
scale dependence that has been the main source of uncertainty at the
NLO.  The central value is shifted downward with respect to all the
previously published NLO results. It is now about $1\sigma$ lower than
the experimental average (\ref{hfag}). The dominant theoretical uncertainty is
currently due to the unknown ${\cal O}\left(\alpha_s \Lambda/m_b\right)$ 
nonperturbative effects. In the two-Higgs-doublet model~II, the
  experimental results favor a charged Higgs boson mass of around 650 GeV.
The 95\%$\,$C.L. bound for this mass amounts to around
295$\,$GeV if all the uncertainties are treated as Gaussian.

We acknowledge support from the 
DFG through SFB/TR~9                            % M.Steinhauser
and a Heisenberg contract,              % T.Hurth
MIUR under Contract No.~2004021808-009, % P.Gambino
the Swiss National Foundation                   % U.Haisch, C.Greub, T.Ewerth
and RTN, BBW-Contract No.~01.0357,      % C.Greub, T.Ewerth
EU-Contracts No.~HPRN-CT-2002-00311     % C.Greub, T.Ewerth, M.Misiak
and  No.~MTRN-CT-2006-035482,   % C.Greub, T.Ewerth, T.Hurth, M.Misiak. M.Steinhauser
Polish KBN grant No. 2~P03B~078~26,     % M.Misiak
the ANSEF~N~05-PS-hepth-0825-338 program,       % H.Asatrian, A.Hovhannisyan, V.Poghosyan
Science and Engineering Research Canada,        % A.Czarnecki
as well as a research fellowship           % A.Mitov 
and the Sofja Kovalevskaja Award                % M.Czakon
of the Alexander von Humboldt Foundation.

{\it Note added}---
Recently, our results from
  Eqs.~(\ref{final}) and (\ref{fit}) were combined in
  Ref.~\cite{Becher:2006pu} with perturbative cutoff-related corrections that
  go beyond a fixed-order calculation \cite{Becher:2006pu,Becher:2005pd}.  
  Because these corrections for $E_0 \leq 1.6\;$GeV do not exceed our
  higher-order uncertainty of $\pm 3\%$, we postpone their consideration
  to a future upgrade of the phenomenological analysis, where other
  contributions of potentially the same size are going to be included,
  too (see Sec. 1 of Ref.~\cite{Asatrian:2006rq}).

\end{document}